\documentclass[a4paper]{article}
\usepackage[preprint]{spconf}
\usepackage{amsmath,graphicx}
\usepackage{multirow}
\usepackage{adjustbox}
\usepackage[hidelinks]{hyperref}
\usepackage{graphicx}
\usepackage{tabularx}
\usepackage{url}
\usepackage{xcolor}
\usepackage{paralist}
\usepackage [english]{babel}
\usepackage [autostyle, english = american]{csquotes}
\MakeOuterQuote{"}

\title{Acoustic-to-articulatory inversion for dysarthric speech: Are pre-trained self-supervised representations favorable?}
\name{Sarthak Kumar Maharana, Krishna Kamal Adidam, Shoumik Nandi, Ajitesh Srivastava}
\address{Ming Hsieh Department of Electrical and Computer Engineering, \\ University of Southern California, Los Angeles, CA, USA - 90087 \\
{\tt\small \{maharana, adidam, shoumikn, ajiteshs\}@usc.edu}}

\begin{document}
\ninept
\maketitle
\begin{abstract}
Acoustic-to-articulatory inversion (AAI) involves mapping from the acoustic to the articulatory space. Signal-processing features like the MFCCs, have been widely used for the AAI task. For subjects with dysarthric speech, AAI is challenging because of an imprecise and indistinct pronunciation. In this work, we perform AAI for dysarthric speech using representations from pre-trained self-supervised learning (SSL) models. We demonstrate the impact of different pre-trained features on this challenging AAI task, at low-resource conditions. In addition, we also condition x-vectors to the extracted SSL features to train a BLSTM network. In the seen case, we experiment with three AAI training schemes (subject-specific, pooled, and fine-tuned). The results, consistent across training schemes, reveal that DeCoAR, in the fine-tuned scheme, achieves a relative improvement of the Pearson Correlation Coefficient (CC) by ${\sim}$1.81\% and ${\sim}$4.56\% for healthy controls and patients, respectively, over MFCCs. We observe similar average trends for different SSL features in the unseen case. Overall, SSL networks like wav2vec, APC, and DeCoAR, trained with feature reconstruction or future timestep prediction tasks, perform well in predicting dysarthric articulatory trajectories.


\end{abstract}
\begin{keywords}
Acoustic-to-articulatory inversion, dysarthria, self-supervised learning, x-vectors, BLSTM.
\end{keywords}

\vspace{-0.15cm}
\section{Introduction}
\label{sec:intro}
\vspace{-0.2cm}
Cerebral Palsy (CP) is a neurological disorder that affects movement, balance, and posture, often caused by damage to the developing brain before or during birth. It is a non-progressive condition. Amyotrophic Lateral Sclerosis (ALS), however, is a progressive neurological disease that leads to muscle weakness and eventual paralysis. Due to a slow muscle response time of a subject, the speech-motor functions are severely affected \cite{kiernan2011amyotrophic}, which eventually leads to weak pronunciation, an affected accent, and unintelligible speech, causing dysarthria \cite{langmore1994physiologic}. As the severity of dysarthria increases, it hurts the movement of articulators, such as the lips, jaw, tongue, and velum, which are responsible for speech production \cite{illa2018comparison, kent1991speech}.


Speech-language pathologists (SLPs) typically employ various speech stimuli, such as reading passages or words, spontaneous speech, or rehearsed speech, for informal evaluations of articulation deterioration \cite{stein2012guide}. To gain a more profound insight into articulation, it is essential to capture the real-time movement of patients' articulators, often achieved through Electromagnetic Articulography (EMA). This data collection approach ensures the simultaneous collection of speech acoustics and articulatory information. Consequently, we are motivated to develop models predicting speech articulator movements based on speech acoustics.



To assist the SLPs and to capture the non-linearity of inversion \cite{toutios2003rough}, deep learning models have been employed. They rely heavily on feature representations to extract meaningful information from the input data and have achieved state-of-the-art performances for AAI \cite{illa2018low, maharana2021acoustic, seneviratne2019multi, udupa2023improved}. Signal-processing features, like the Mel-frequency Cepstral Coefficients (MFCCs), have been shown to be optimal for AAI \cite{ghosh2010generalized}. However, there's a rising demand for using a parameterized representation of speech acoustics, as input features, learned via self-supervised learning (SSL) methods \cite{baevski2019vq, liu2020mockingjay}. In \cite{udupa2023improved}, Udupa et al. extensively studied the effect of different pre-trained SSL features on AAI tasks. They proved that a better utterance level generalizability was obtained using SSL features, making it a promising choice over MFCCs.



To the best of our knowledge, there has been no work extending SSL for AAI on \underline{dysarthric} speech. In this study, we wish to explore a variety of SSL models such as APC \cite{chung2019unsupervised}, NPC \cite{liu2020non}, DeCoAR \cite{ling2020deep}, wav2vec \cite{schneider2019wav2vec}, TERA \cite{Liu_2021}, vq$\_$wav2vec \cite{baevski2019vq}, and Mockingjay \cite{liu2020mockingjay} that have their respective training schemes, which result in different model parameters used to extract features that would then be fed to a sequential model like the bidirectional LSTM (BLSTM), to estimate the articulatory trajectories \cite{illa2018low,maharana2021acoustic}. We hypothesize that learning a robust and rich representation space for dysarthric speakers via pre-trained SSL models is important. This would capture the speaking styles, different speaking rates \cite{illa2020impact}, and speaker identities \cite{illa2020closed}, which would aid in generalizing to unknown dysarthric speakers, to boost the AAI performance. In addition, to preserve speaker-specific information, we plan on conditioning the obtained SSL features with their corresponding x-vector embeddings \cite{snyder2018x} to learn rich acoustic-articulatory mappings of multiple speakers. 


\textbf{Key contributions} - Below, we articulate the key contributions of our work at different levels:
\begin{inparaitem}
    \item We propose the first work demonstrating the effects of different SSL features for AAI of dysarthric speech at low-resource data conditions. In this study, we explore if using pre-trained predictive and contrastive SSL models benefits AAI for dysarthric speech in seen and unseen subject conditions.
    \item We illustrate and analyze articulatory level performances by the best SSL feature against baseline MFCCs, in the seen subject conditions.
    \item We empirically show the benefits of speaker-specific embeddings like the x-vectors \cite{snyder2018x} to boost AAI performance.
\end{inparaitem}
\vspace{-0.4cm}
\section{Dataset}
\label{sec:dataset}
\vspace{-0.2cm}
In this work, we use the TORGO dataset \cite{rudzicz2012torgo} that comprises aligned speech acoustics and measured 3D EMA articulatory trajectories from 15 speakers with either CP or ALS and matched healthy controls. Out of the acoustic-articulatory data available from 7 speakers only, we choose 4 speakers with \underline{complete} parallel acoustic-articulatory data i.e., 2 male healthy controls (MC01 and MC04) and 2 female patients (F03 and F04). Articulatory movements, in the X and Y directions, from sensor coils attached to the tongue tip (TT), tongue middle (TM), tongue back (TB), jaw (JAW), lower lip (LL), and upper lip (UL) are considered. The tongue's middle and back portions are also called the body and dorsum, respectively. We denote TM and TB as TB and TD, henceforth. This results in 12-dimensional articulatory feature vectors denoted as UL$_x$, UL$_y$, LL$_x$, LL$_y$, JAW$_x$, JAW$_y$, TT$_x$, TT$_y$, TB$_x$, TB$_y$, TD$_x$, and TD$_y$. Illa et al. \cite{illa2018comparison} proposed the addition of kinematic features like velocity and acceleration to the articulatory features to improve speech-based classification between ALS patients and healthy controls. This results in a 6-dimensional velocity and acceleration articulatory feature vector along the X and Y directions, each. Finally, we obtain 24-dimensional articulatory feature vectors upon concatenation.

Table \ref{tab:durdata} provides the speech utterance duration for healthy controls and patients used in this work. Row "Severity" indicates the amount of severity of dysarthria for both patients. Row "Number of utterances" reports the total number of utterances uttered by each subject, and "Total duration" indicates the total duration of speech segments, in seconds, uttered by each subject. Since collecting articulatory data is difficult because of the discomfort caused by sensor coils, the total duration and number of utterances for patients were lesser than those of healthy controls. On average, 7535 utterances from healthy controls and 3066 utterances from each dysarthric speaker were recorded, performing speech tasks from the TIMIT database \cite{zue1990speech}. Silence and noise from each sentence were later removed, referring to the provided transcriptions.

\begin{table}[htb!]
\centering
\resizebox{0.32\textwidth}{!}{%
\begin{tabular}{|c|cc|cc|}
\hline
\multirow{2}{*}{\textbf{Speakers}}                              & \multicolumn{2}{c|}{\textbf{Healthy Controls}} & \multicolumn{2}{c|}{\textbf{Patients}} \\ \cline{2-5} 
         & \multicolumn{1}{c|}{\textbf{MC01}} & \textbf{MC04} & \multicolumn{1}{c|}{\textbf{F03}} & \textbf{F04} \\ \hline
Severity & \multicolumn{1}{c|}{-}             & -             & \multicolumn{1}{c|}{Moderate}     & Mild         \\ \hline
\begin{tabular}[c]{@{}c@{}}Number of \\ utterances\end{tabular} & \multicolumn{1}{c|}{8186}         & 6884       & \multicolumn{1}{c|}{3534}    & 2599    \\ \hline
\begin{tabular}[c]{@{}c@{}}Total \\ duration\end{tabular}       & \multicolumn{1}{c|}{1032.33}    & 774.97   & \multicolumn{1}{c|}{385.57} & 298.96 \\ \hline
\end{tabular}%
}
\caption{Speech duration (in seconds) for the 2 male healthy controls and the 2 female dysarthric patients constituting the TORGO dataset.}
\label{tab:durdata}
\end{table}

\vspace{-0.8cm}
\section{Proposed Methodology}
\vspace{-0.25cm}
Acoustic-to-articulatory inversion (AAI) is a regression task, in which the one-to-many mapping from the acoustic space onto the articulatory space, is non-linear, complex, and ill-posed \cite{kirchho1999robust,wu2015acoustic}. Hence, various approaches have been proposed in the state-of-the-art literature to model this complexity and non-linearity through neural networks like BLSTMs \cite{illa2018low, maharana2021acoustic} and Transformers \cite{udupa2021estimating}. However, in this work, we employ the BLSTM architecture, inspired from \cite{illa2018low, maharana2021acoustic}, to perform AAI. Since the amount of acoustic-articulatory data from healthy controls and dysarthric patients might be insufficient, we believe that learning a rich and robust representation space via pre-trained SSL models could instead be beneficial to perform AAI on dysarthric speech over transfer learning and joint training using a rich yet mismatched cross-corpus \cite{maharana2021acoustic}.

\begin{figure}[htb!]
 \centering
  \includegraphics[width=0.42\textwidth]{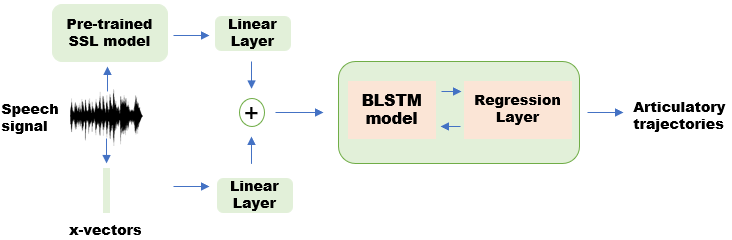}
      \caption{Block diagram of the proposed AAI model with pre-trained SSL features as the input acoustic features, conditioned with x-vectors.}
    \label{proposed}
\end{figure}

This work proposes a novel approach that employs using features from pre-trained SSL models for dysarthric AAI particularly, i.e., it combines the usage of features from the pre-trained SSL models, as acoustic features \cite{udupa2023improved} and speaker-specific information via x-vectors \cite{maharana2021acoustic}. Fig. \ref{proposed} illustrates our block diagram, adopted from \cite{illa2018low, maharana2021acoustic}, for this work.

As a baseline acoustic feature, we consider MFCCs. These features have been conventionally shown to be optimal in previous AAI tasks \cite{ghosh2010generalized, illa2018low}. For pre-trained SSL features, based on the training schemes of the respective models, these features are extracted from models trained on their respective predictive or contrastive loss functions. It is important to note that the weights of the SSL models are not trainable. The feature representations are extracted from the SSL model, by giving the speech signal as the input.

This novel idea of using pre-trained SSL features, along with conditioning with x-vectors \cite{snyder2018x}, will aid in learning better representations for dysarthric speech. This, in turn, will help in the better prediction of dysarthric articulatory trajectories. 

\vspace{-0.7cm}

\section{Experimental Setup}
\begin{figure}[htb!]
\centering
\includegraphics[trim = 0mm 0mm 1mm 1mm, clip, width = 0.3\textwidth]{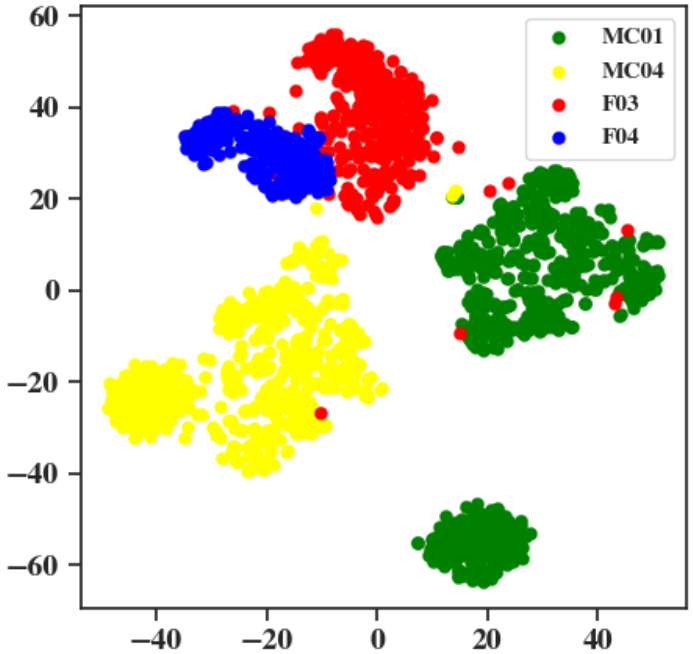}
  \caption{t-SNE plot of x-vector speaker embeddings of healthy controls (MC01, MC04) and patients (F03, F04), used in this work.}
\label{xvecs}
\end{figure}

\vspace{-0.2cm}
\emph{\underline{Feature extraction}:} As an initial step, we pre-process the recorded speech waveforms and their corresponding articulatory features. The speech waveforms are downsampled to 16 kHz. The articulatory features are downsampled to 100 Hz, from the initial 200 Hz, and low-pass filtered at a cutoff frequency of 25 Hz. This was done due to the presence of high-frequency noise that could be incurred during the recording session. We use the Kaldi toolkit \cite{povey2011kaldi} to compute MFCCs, using a window length of 25ms and a shift of 10ms. This way, the MFCCs have a one-to-one correspondence with the articulatory trajectories. For SSL feature extraction, we use the upstream model weights from s3prl \footnote{\href{https://github.com/s3prl/s3prl}{\scriptsize{https://github.com/s3prl/s3prl}}}, for the initialization of each of the SSL models. We initialize using the default pre-trained weights, as provided by s3prl. Mean and variance normalization is performed at an utterance level, across each dimension of the acoustic and articulatory features. The SSL feature representations are derived from the final layer of the SSL model \cite{udupa2023improved}. For the computation of x-vectors, we use the Kaldi toolkit \cite{povey2011kaldi}, using a pre-trained model trained on the VoxCeleb database \cite{nagrani2020voxceleb}. Fig. \ref{xvecs} illustrates the x-vector speaker embeddings, for each subject, after t-SNE \cite{van2008visualizing}. We observe that the embeddings obtained from the pre-trained model are able to discriminate between the speakers. However, an interesting observation is that the x-vectors for certain sentences from F03 overlap with those of healthy controls. This could be due to the low dysarthric severity of F03 which translates to similar speaking rates or styles. In addition, we also observe that the x-vectors of MC01 form two separate clusters. This could be due to different speaking rates and lexical patterns across sentences for MC01.

\emph{\underline{Training schemes and network parameters}:} We perform experiments in two subject conditions - seen and unseen. In the seen conditions, for each subject, we uniformly split the sentences as 90\% for training and the remaining 10\% for testing. We make sure that the sentences do not overlap across the two sets. Here, the AAI models are trained in three training schemes:
\begin{inparaitem}
    \item Subject-specific: The AAI model is trained on the training data of each subject and tested on its corresponding test set.
    \item Pooled: A single AAI model is trained on a combined set of training data of all subjects. During inference, the test sets of the subjects are aggregated.
    \item Fine-tuned: Trained parameters from the pooled scheme are initialized and fine-tuned on the training data of each subject and re-trained.
\end{inparaitem}
In the unseen subject conditions, we employ the leave-one-person-out approach i.e., we train an AAI model on the data of all subjects except the one chosen as the test subject. We test the AAI model on the subject that is left out. In both the training conditions (seen and unseen), we consider a standard 5-fold cross-validation setup out of which 4 folds are considered for training and the other fold is considered for validation, in a round-robin fashion.

Due to variable sequence lengths in a batch of data, we performed zero-padding based on the length of the maximum sequence in that particular batch \cite{udupa2023improved}. We randomly initialized the parameters of the 3-layered BLSTM network, where each layer uses 256 hidden LSTM units. For the acoustic feature, the input dense layer is initialized with 200 units. The speaker embedding dense layer, for the x-vectors, is 32 units. A batch size of 5 is used for training over 50 epochs, with a learning rate of $10^{-4}$, and a weight decay of $10^{-6}$. We employ a learning rate scheduler to avoid the stagnation of learning. Network parameters are optimized using the Adam optimizer, where the regression estimation of articulatory trajectories is minimized using a mean-squared error (MSE) loss function. Based on the validation loss, early stopping is also performed. We use PyTorch \cite{paszke2019pytorch} to train all our AAI models.

\emph{\underline{Evaluation metrics}:} We use the Pearson Correlation Coefficient (CC) \cite{illa2018low, ghosh2010generalized} between the ground-truth and the predicted articulatory trajectories. The first 12 dimensions of the trajectories are considered since they correspond to the raw positions of the six articulators, along the X and Y directions, as considered in this work.



\vspace{-0.3cm}
\section{Results and Discussions}
\vspace{-0.2cm}
\subsection{Effect of speaker-specific embeddings}
\vspace{-0.2cm}
Table \ref{tab:LSTM} reports the average CC (standard deviation) on the test set, across all the five folds, articulators, and sentences, in the pooled training seen subject scheme, for different speaker-specific embeddings conditioned on MFCCs. To verify such an impact on the acoustic features, we investigate the performance by conditioning the MFCCs with its statistics i.e., its mean and standard deviation, which results in a dimension that is double the input dimension of MFCCs. We also condition the MFCCs with their corresponding x-vectors. The results reveal that the AAI model (256 LSTM units) with input MFCCs and conditioned with x-vectors outperforms "only MFCCs" and MFCCs with statistics by ${\sim}$4.52\% and ${\sim}$2.99\% for patients respectively, and, ${\sim}$1.09\% and ${\sim}$1.06\% for healthy controls respectively. Hence, we report the remaining results where the acoustic features (MFCCs or pre-trained SSL features) are conditioned on its x-vectors.

Since the amount of training data might be insufficient, the proposed network may have a chance of overfitting. So, we experiment with different LSTM units for each layer. The performances of the different configurations of MFCCs improve with an increase in the LSTM units. Due to an increase in network complexity and computation cost \cite{illa2018low}, we use and report the results on a BLSTM network with 256 LSTM units in each layer for the remainder of our work.


\begin{table}[htb!]
\centering
\resizebox{0.4\textwidth}{!}{%
\begin{tabular}{|c|cc|cc|cc|}
\hline
\multirow{2}{*}{\begin{tabular}[c]{@{}c@{}}LSTM\\ units\end{tabular}} &
  \multicolumn{2}{c|}{MFCCs} &
  \multicolumn{2}{c|}{\begin{tabular}[c]{@{}c@{}}MFCCs\\ w/ statistics\end{tabular}} &
  \multicolumn{2}{c|}{\begin{tabular}[c]{@{}c@{}}MFCCs\\ w/ x-vectors\end{tabular}} \\ \cline{2-7} 
 &
  \multicolumn{1}{c|}{\begin{tabular}[c]{@{}c@{}}Healthy\\ Controls\end{tabular}} &
  Patients &
  \multicolumn{1}{c|}{\begin{tabular}[c]{@{}c@{}}Healthy\\ Controls\end{tabular}} &
  Patients &
  \multicolumn{1}{c|}{\begin{tabular}[c]{@{}c@{}}Healthy\\ Controls\end{tabular}} &
  Patients \\ \hline
32 &
  \multicolumn{1}{c|}{\begin{tabular}[c]{@{}c@{}}0.646\\ (0.0667)\end{tabular}} &
  \begin{tabular}[c]{@{}c@{}}0.4505\\ (0.1328)\end{tabular} &
  \multicolumn{1}{c|}{\begin{tabular}[c]{@{}c@{}}0.6379\\ (0.075)\end{tabular}} &
  \begin{tabular}[c]{@{}c@{}}0.4633\\ (0.1352)\end{tabular} &
  \multicolumn{1}{c|}{\begin{tabular}[c]{@{}c@{}}0.6532\\ (0.0677)\end{tabular}} &
  \begin{tabular}[c]{@{}c@{}}0.4747\\ (0.1308)\end{tabular} \\ \hline
64 &
  \multicolumn{1}{c|}{\begin{tabular}[c]{@{}c@{}}0.6748\\ (0.0649)\end{tabular}} &
  \begin{tabular}[c]{@{}c@{}}0.4782\\ (0.1233)\end{tabular} &
  \multicolumn{1}{c|}{\begin{tabular}[c]{@{}c@{}}0.6771\\ (0.064)\end{tabular}} &
  \begin{tabular}[c]{@{}c@{}}0.4895\\ (0.1283)\end{tabular} &
  \multicolumn{1}{c|}{\begin{tabular}[c]{@{}c@{}}0.6895\\ (0.0637)\end{tabular}} &
  \begin{tabular}[c]{@{}c@{}}0.496\\ (0.1251)\end{tabular} \\ \hline
128 &
  \multicolumn{1}{c|}{\begin{tabular}[c]{@{}c@{}}0.7083\\ (0.0581)\end{tabular}} &
  \begin{tabular}[c]{@{}c@{}}0.504\\ (0.1152)\end{tabular} &
  \multicolumn{1}{c|}{\begin{tabular}[c]{@{}c@{}}0.7069\\ (0.0575)\end{tabular}} &
  \begin{tabular}[c]{@{}c@{}}0.5086\\ (0.1157)\end{tabular} &
  \multicolumn{1}{c|}{\begin{tabular}[c]{@{}c@{}}0.7224\\ (0.0536)\end{tabular}} &
  \begin{tabular}[c]{@{}c@{}}0.5188\\ (0.1232)\end{tabular} \\ \hline
256 &
  \multicolumn{1}{c|}{\begin{tabular}[c]{@{}c@{}}0.7412\\ (0.0488)\end{tabular}} &
  \begin{tabular}[c]{@{}c@{}}0.5201\\ (0.1073)\end{tabular} &
  \multicolumn{1}{c|}{\begin{tabular}[c]{@{}c@{}}0.7414\\ (0.0487)\end{tabular}} &
  \begin{tabular}[c]{@{}c@{}}0.5278\\ (0.1234)\end{tabular} &
  \multicolumn{1}{c|}{\begin{tabular}[c]{@{}c@{}}\textbf{0.7493}\\ \textbf{(0.048)}\end{tabular}} &
  \begin{tabular}[c]{@{}c@{}}\textbf{0.5436}\\ \textbf{(0.1194)}\end{tabular} \\ \hline
\end{tabular}%
}
\caption{Average CC (standard deviation) across the test sets in the pooled training seen subject scheme, using only MFCCs, MFCCs conditioned with statistics, and MFCCs conditioned with x-vectors, for different LSTM units. We average across all the articulators, sentences, and folds.}
\label{tab:LSTM}
\end{table}

\vspace{-0.4cm}
\begin{table}[htb!]
\centering
\resizebox{0.45\textwidth}{!}{%
\begin{tabular}{|c|cc|cc|cc|}
\hline
\multirow{2}{*}{\textbf{Features}} &
  \multicolumn{2}{c|}{\textbf{Subject-specific}} &
  \multicolumn{2}{c|}{\textbf{Pooled}} &
  \multicolumn{2}{c|}{\textbf{Fine-tuned}} \\ \cline{2-7} 
 &
  \multicolumn{1}{c|}{\textbf{\begin{tabular}[c]{@{}c@{}}Healthy\\ Controls\end{tabular}}} &
  \textbf{Patients} &
  \multicolumn{1}{c|}{\textbf{\begin{tabular}[c]{@{}c@{}}Healthy\\ Controls\end{tabular}}} &
  \textbf{Patients} &
  \multicolumn{1}{c|}{\textbf{\begin{tabular}[c]{@{}c@{}}Healthy\\ Controls\end{tabular}}} &
  \textbf{Patients} \\ \hline
MFCCs &
  \multicolumn{1}{c|}{\begin{tabular}[c]{@{}c@{}}0.7627\\ (0.0578)\end{tabular}} &
  \begin{tabular}[c]{@{}c@{}}0.5534\\ (0.1374)\end{tabular} &
  \multicolumn{1}{c|}{\begin{tabular}[c]{@{}c@{}}0.7493\\ (0.048)\end{tabular}} &
  \begin{tabular}[c]{@{}c@{}}0.5436\\ (0.1194)\end{tabular} &
  \multicolumn{1}{c|}{\begin{tabular}[c]{@{}c@{}}0.7629\\ (0.0572)\end{tabular}} &
  \begin{tabular}[c]{@{}c@{}}0.5808\\ (0.1200)\end{tabular} \\ \hline
wav2vec &
  \multicolumn{1}{c|}{\begin{tabular}[c]{@{}c@{}}0.7648\\ (0.0561)\end{tabular}} &
  \begin{tabular}[c]{@{}c@{}}0.5591\\ (0.1283)\end{tabular} &
  \multicolumn{1}{c|}{\begin{tabular}[c]{@{}c@{}}0.756\\ (0.0437)\end{tabular}} &
  \begin{tabular}[c]{@{}c@{}}0.5808\\ (0.1106)\end{tabular} &
  \multicolumn{1}{c|}{\begin{tabular}[c]{@{}c@{}}0.7649\\ (0.0554)\end{tabular}} &
  \begin{tabular}[c]{@{}c@{}}0.593\\ (0.1216)\end{tabular} \\ \hline
APC &
  \multicolumn{1}{c|}{\begin{tabular}[c]{@{}c@{}}0.7544\\ (0.0596)\end{tabular}} &
  \begin{tabular}[c]{@{}c@{}}0.5438\\ (0.1265)\end{tabular} &
  \multicolumn{1}{c|}{\begin{tabular}[c]{@{}c@{}}0.7481\\ (0.0458)\end{tabular}} &
  \begin{tabular}[c]{@{}c@{}}0.5717\\ (0.1159)\end{tabular} &
  \multicolumn{1}{c|}{\begin{tabular}[c]{@{}c@{}}0.7642\\ (0.0551)\end{tabular}} &
  \begin{tabular}[c]{@{}c@{}}0.5867\\ (0.1224)\end{tabular} \\ \hline
NPC &
  \multicolumn{1}{c|}{\begin{tabular}[c]{@{}c@{}}0.7561\\ (0.059)\end{tabular}} &
  \begin{tabular}[c]{@{}c@{}}0.5441\\ (0.1443)\end{tabular} &
  \multicolumn{1}{c|}{\begin{tabular}[c]{@{}c@{}}0.7501\\ (0.0468)\end{tabular}} &
  \begin{tabular}[c]{@{}c@{}}0.5421\\ (0.1116)\end{tabular} &
  \multicolumn{1}{c|}{\begin{tabular}[c]{@{}c@{}}0.7592\\ (0.0594)\end{tabular}} &
  \begin{tabular}[c]{@{}c@{}}0.5563\\ (0.1357)\end{tabular} \\ \hline
DeCoAR &
  \multicolumn{1}{c|}{\textbf{\begin{tabular}[c]{@{}c@{}}0.7699\\ (0.0540)\end{tabular}}} &
  \textbf{\begin{tabular}[c]{@{}c@{}}0.5832\\ (0.1344)\end{tabular}} &
  \multicolumn{1}{c|}{\textbf{\begin{tabular}[c]{@{}c@{}}0.7628\\ (0.0445)\end{tabular}}} &
  \textbf{\begin{tabular}[c]{@{}c@{}}0.5928\\ (0.1187)\end{tabular}} &
  \multicolumn{1}{c|}{\textbf{\begin{tabular}[c]{@{}c@{}}0.7767\\ (0.0562)\end{tabular}}} &
  \textbf{\begin{tabular}[c]{@{}c@{}}0.6073\\ (0.1312)\end{tabular}} \\ \hline
TERA &
  \multicolumn{1}{c|}{\begin{tabular}[c]{@{}c@{}}0.7481\\ (0.0613)\end{tabular}} &
  \begin{tabular}[c]{@{}c@{}}0.5451\\ (0.1322)\end{tabular} &
  \multicolumn{1}{c|}{\begin{tabular}[c]{@{}c@{}}0.7515\\ (0.0492)\end{tabular}} &
  \begin{tabular}[c]{@{}c@{}}0.5562\\ (0.1242)\end{tabular} &
  \multicolumn{1}{c|}{\begin{tabular}[c]{@{}c@{}}0.7657\\ (0.0587)\end{tabular}} &
  \begin{tabular}[c]{@{}c@{}}0.5702\\ (0.1348)\end{tabular} \\ \hline
Mockingjay &
  \multicolumn{1}{c|}{\begin{tabular}[c]{@{}c@{}}0.7291\\ (0.0624)\end{tabular}} &
  \begin{tabular}[c]{@{}c@{}}0.5121\\ (0.1437)\end{tabular} &
  \multicolumn{1}{c|}{\begin{tabular}[c]{@{}c@{}}0.7289\\ (0.0499)\end{tabular}} &
  \begin{tabular}[c]{@{}c@{}}0.5287\\ (0.118)\end{tabular} &
  \multicolumn{1}{c|}{\begin{tabular}[c]{@{}c@{}}0.7428\\ (0.0595)\end{tabular}} &
  \begin{tabular}[c]{@{}c@{}}0.547\\ (0.1326)\end{tabular} \\ \hline
vq\textunderscore wav2vec &
  \multicolumn{1}{c|}{\begin{tabular}[c]{@{}c@{}}0.7192\\ (0.0683)\end{tabular}} &
  \begin{tabular}[c]{@{}c@{}}0.5299\\ (0.1337)\end{tabular} &
  \multicolumn{1}{c|}{\begin{tabular}[c]{@{}c@{}}0.7209\\ (0.0559)\end{tabular}} &
  \begin{tabular}[c]{@{}c@{}}0.5631\\ (0.1066)\end{tabular} &
  \multicolumn{1}{c|}{\begin{tabular}[c]{@{}c@{}}0.7361\\ (0.0631)\end{tabular}} &
  \begin{tabular}[c]{@{}c@{}}0.5824\\ (0.1154)\end{tabular} \\ \hline
\end{tabular}%
}
\caption{Average CC (standard deviation) in the seen conditions across the respective test sets in the subject-specific, pooled, and fine-tuned training schemes, for different SSL input features. We report results averaged across all the articulators, sentences, and folds.}
\label{tab:ssl}
\end{table}

\vspace{-0.75cm}
\subsection{Impact of different pre-trained SSL features}
\vspace{-0.2cm}
\textit{\underline{Seen subject evaluation}:} Table \ref{tab:ssl} reports the average CC (standard deviation), in the seen subject conditions, across the respective test sets, averaged across all the articulators, sentences, and folds, for all the training schemes i.e., subject-specific, pooled, and fine-tuned, for the different SSL features, in this work, as the input. 
In the subject-specific scheme, we notice that features from DeCoAR perform better over MFCCs with relative CC improvements of ${\sim}$0.94\% and ${\sim}$5.34\% over MFCCs, for healthy controls and patients respectively. It is interesting to note that vq\_wav2vec, Mockingjay, and TERA, perform marginally poorly, but APC and NPC achieve comparable performances to MFCCs. The drop in performance by vq\_wav2vec could be due to the quantization step in its training objective \cite{baevski2019vq}. TERA and Mockingjay which are trained with a masked loss function to predict the sequence of words in a sentence, seem to give a lower performance \cite{liu2020mockingjay, Liu_2021} over here. In addition, a stronger reason could be that in the subject-specific scheme, the amount of training data from each subject is less, contributing to overfitting and poor generalization on the corresponding test set. For the pooled and fine-tuned training schemes, we observe that the performances of all the SSL features are better than MFCCs, for both healthy control and patients, except for Mockingjay and vq\_wav2vec. However, vq\_wav2vec performs better than MFCCs for patients. To be specific, DeCoAR performs the best across all the setups with CC improvements of ${\sim}$1.81\% and ${\sim}$4.56\% for healthy controls and patients respectively, in the fine-tuned case. The combination of unsupervised pre-training and supervised fine-tuning, in DeCoAR, allows the network to learn general features of speech and task-specific features \cite{ling2020deep} i.e., articulatory trajectories of healthy controls and patients. This validates that SSL pre-trained features are more suited and robust acoustic features for dysarthric AAI. Interestingly, on the whole, the results using MFCCs for patients are competitive due to lower dysarthria severity.

\textit{\underline{Unseen subject evaluation}:} Table \ref{tab:unseen} reports the average CC (standard deviation), in the unseen subject conditions, across all the sentences and articulators, for each subject. We observe similar average performances, as in the seen conditions, across the different subjects. In particular, wav2vec and DeCoAR perform better over MFCCs, for all the subjects. For F03 (with moderate dysarthric severity), the CC can be improved from 0.4201 to 0.4659, using DeCoAR. It is to be noted that AAI on unseen dysarthric subjects is challenging. Pre-trained SSL features combined with the conditioning of x-vectors led to a better utterance level generalisability on subjects unseen during training \cite{illa2020speaker}. This confirms the effectiveness of pre-trained SSL features in performing unseen dysarthric AAI.

\vspace{-0.2cm}
\begin{table}[htb!]
\centering
\resizebox{0.3\textwidth}{!}{%
\begin{tabular}{|c|cccc|}
\hline
\multirow{2}{*}{\textbf{Features}} &
  \multicolumn{4}{c|}{\textbf{Unseen subjects}} \\ \cline{2-5} 
 &
  \multicolumn{1}{c|}{\textbf{F03}} &
  \multicolumn{1}{c|}{\textbf{F04}} &
  \multicolumn{1}{c|}{\textbf{MC01}} &
  \textbf{MC04} \\ \hline
MFCCs &
  \multicolumn{1}{c|}{\begin{tabular}[c]{@{}c@{}}0.4201\\ (0.1342)\end{tabular}} &
  \multicolumn{1}{c|}{\begin{tabular}[c]{@{}c@{}}0.4505\\ (0.187)\end{tabular}} &
  \multicolumn{1}{c|}{\begin{tabular}[c]{@{}c@{}}0.4439\\ (0.0861)\end{tabular}} &
  \begin{tabular}[c]{@{}c@{}}0.568\\ (0.1059)\end{tabular} \\ \hline
wav2vec &
  \multicolumn{1}{c|}{\begin{tabular}[c]{@{}c@{}}0.4422\\ (0.1362)\end{tabular}} &
  \multicolumn{1}{c|}{\begin{tabular}[c]{@{}c@{}}\textbf{0.5126}\\ \textbf{(0.2003)}\end{tabular}} &
  \multicolumn{1}{c|}{\begin{tabular}[c]{@{}c@{}}\textbf{0.5134}\\ \textbf{(0.0879)}\end{tabular}} &
  \begin{tabular}[c]{@{}c@{}}0.5781\\ (0.1179)\end{tabular} \\ \hline
APC &
  \multicolumn{1}{c|}{\begin{tabular}[c]{@{}c@{}}0.4284\\ (0.1489)\end{tabular}} &
  \multicolumn{1}{c|}{\begin{tabular}[c]{@{}c@{}}0.4502\\ (0.1826)\end{tabular}} &
  \multicolumn{1}{c|}{\begin{tabular}[c]{@{}c@{}}0.4879\\ (0.0875)\end{tabular}} &
  \begin{tabular}[c]{@{}c@{}}0.5608\\ (0.1111)\end{tabular} \\ \hline
NPC &
  \multicolumn{1}{c|}{\begin{tabular}[c]{@{}c@{}}0.4255\\ (0.1427)\end{tabular}} &
  \multicolumn{1}{c|}{\begin{tabular}[c]{@{}c@{}}0.4257\\ (0.2106)\end{tabular}} &
  \multicolumn{1}{c|}{\begin{tabular}[c]{@{}c@{}}0.4852\\ (0.0859)\end{tabular}} &
  \begin{tabular}[c]{@{}c@{}}0.5458\\ (0.1237)\end{tabular} \\ \hline
DeCoAR &
  \multicolumn{1}{c|}{\begin{tabular}[c]{@{}c@{}}\textbf{0.4659}\\ \textbf{(0.1414)}\end{tabular}} &
  \multicolumn{1}{c|}{\begin{tabular}[c]{@{}c@{}}0.4953\\ (0.1931)\end{tabular}} &
  \multicolumn{1}{c|}{\begin{tabular}[c]{@{}c@{}}0.5102\\ (0.0863)\end{tabular}} &
  \begin{tabular}[c]{@{}c@{}}\textbf{0.5938}\\ \textbf{(0.1022)}\end{tabular} \\ \hline
TERA &
  \multicolumn{1}{c|}{\begin{tabular}[c]{@{}c@{}}0.4437\\ (0.1418)\end{tabular}} &
  \multicolumn{1}{c|}{\begin{tabular}[c]{@{}c@{}}0.4471\\ (0.191)\end{tabular}} &
  \multicolumn{1}{c|}{\begin{tabular}[c]{@{}c@{}}0.4901\\ (0.0937)\end{tabular}} &
  \begin{tabular}[c]{@{}c@{}}0.5504\\ (0.1358)\end{tabular} \\ \hline
Mockingjay &
  \multicolumn{1}{c|}{\begin{tabular}[c]{@{}c@{}}0.4182\\ (0.1329)\end{tabular}} &
  \multicolumn{1}{c|}{\begin{tabular}[c]{@{}c@{}}0.4139\\ (0.1807)\end{tabular}} &
  \multicolumn{1}{c|}{\begin{tabular}[c]{@{}c@{}}0.4503\\ (0.0923)\end{tabular}} &
  \begin{tabular}[c]{@{}c@{}}0.5031\\ (0.1407)\end{tabular} \\ \hline
vq\_wav2vec &
  \multicolumn{1}{c|}{\begin{tabular}[c]{@{}c@{}}0.4341\\ (0.1432)\end{tabular}} &
  \multicolumn{1}{c|}{\begin{tabular}[c]{@{}c@{}}0.4823\\ (0.193)\end{tabular}} &
  \multicolumn{1}{c|}{\begin{tabular}[c]{@{}c@{}}0.4836\\ (0.0994)\end{tabular}} &
   \begin{tabular}[c]{@{}c@{}}0.5874\\ (0.1149)\end{tabular} \\ \hline
\end{tabular}%
}
\caption{Average CC (standard deviation) in the unseen conditions, across all the sentences and articulators for different SSL input features. The results are reported for each unseen subject during training of the AAI model. }
\label{tab:unseen}
\end{table}

\vspace{-0.5cm}

\begin{figure}[htb!]
 \centering
  \includegraphics[width=0.4\textwidth]{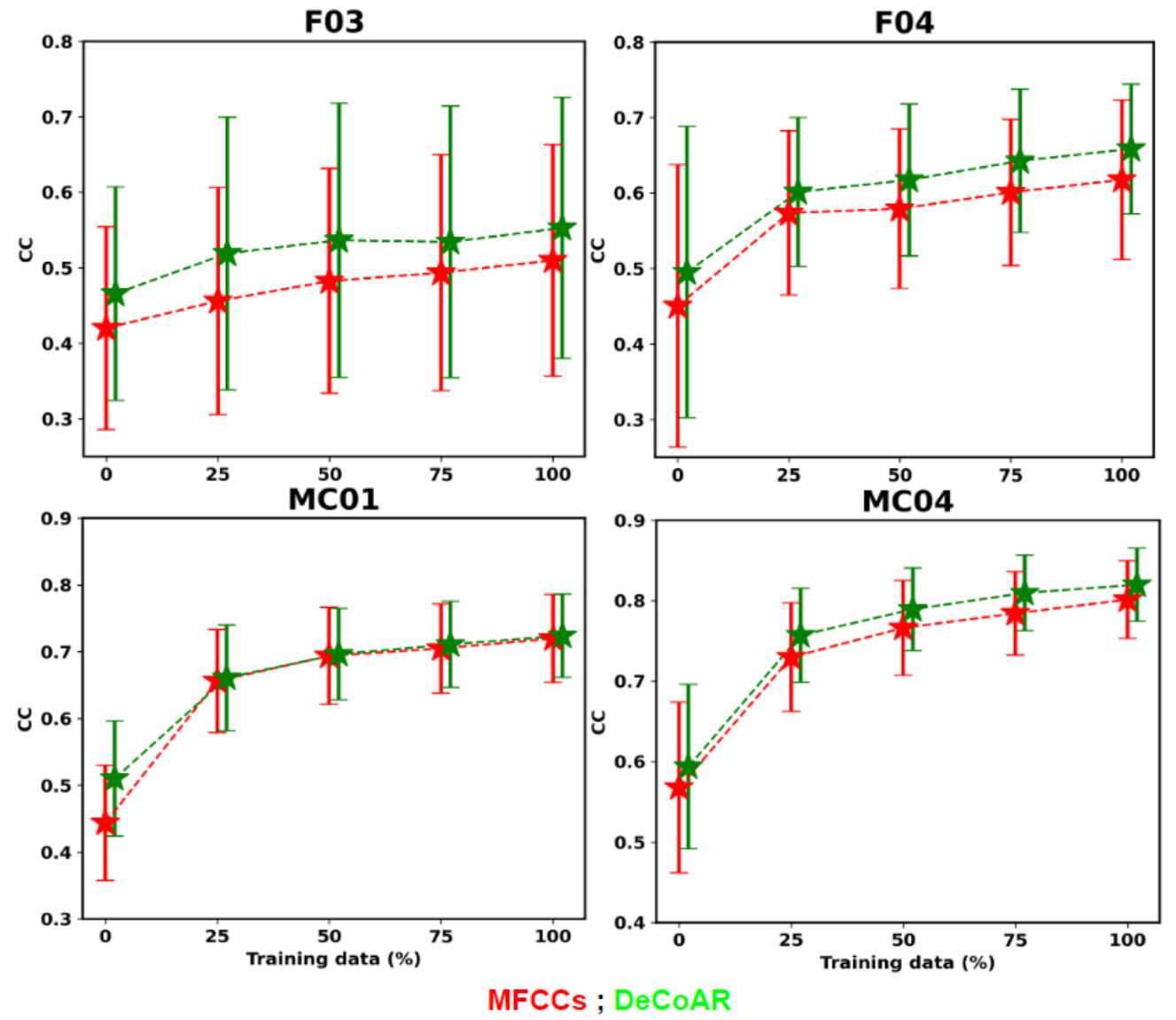}
      \caption{Average CC (with error bars) for unseen target subjects using \textcolor{red}{MFCCs} and \textcolor{green}{DeCoAR} at varying amounts of training data (\textit{t\%}), averaged across all the articulators, fold, and sentences. \textit{t} = 0\ refers to the unseen subject evaluation. }
    \label{p_data}
\end{figure}

\vspace{0.1cm}
\textit{\underline{Effect of data adaptation}:} Here, we study the amount of training data required from an unseen target subject to achieve a comparable performance using the same subject's complete training data. We begin by selecting \textit{t\%} of training data from a subject and fine-tuning the AAI model weights that have been trained on the remaining reference subjects. In all the plots of Fig. \ref{p_data}, the x-axis reports the various \textit{t\%} of training data that we choose, and the y-axis reports the average CC across all the articulators, sentences, and folds. Clearly, \textit{t}=0 refers to an unseen subject evaluation. From Fig. \ref{p_data}, we observe that for all the subjects with \textit{t}$\ge$0, DeCoAR performs better than MFCCs. The better performances of DeCoAR by using small amounts of training data from a target subject could be attributed to the rich representation space and multiple acoustic-articulatory mappings it learns, for the dysarthric patients and healthy controls. For F03 and F04 at the \textit{t}=50 mark, DeCoAR archives relative improvements of ${\sim}$11.23\% and ${\sim}$6.59\% respectively, over MFCCs. Overall, experimental results suggest that using small amounts of training data (instead of full training, \textit{t}=100) with pre-trained SSL features, from an unseen target subject, is beneficial for dysarthric AAI since collecting full acoustic-articulatory data from a subject is tedious.

\vspace{-0.7cm}
\subsection{Articulatory specific results}
\begin{figure}[htb!]
 \centering
  \includegraphics[width=0.45\textwidth]{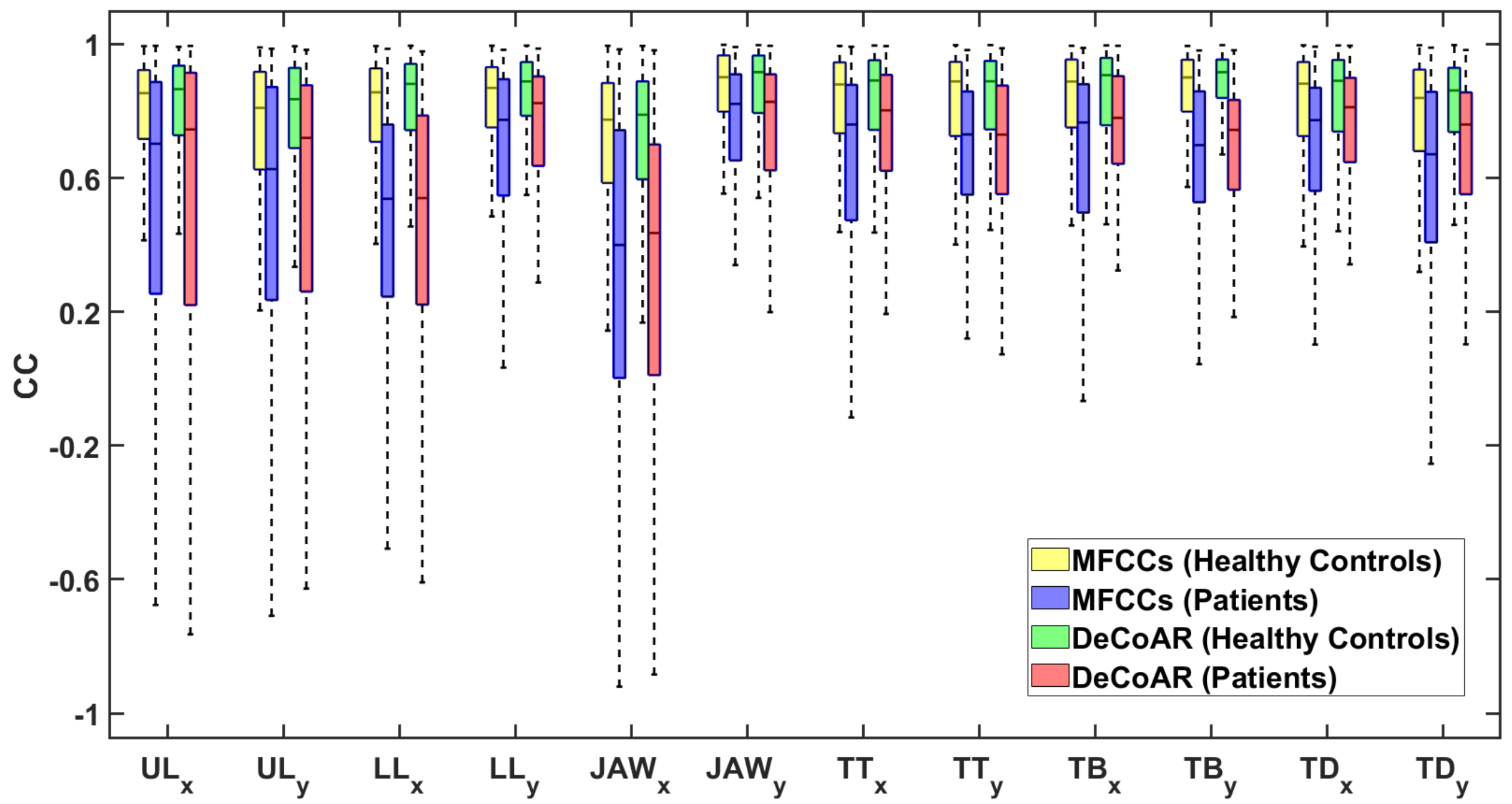}
      \caption{CC values for each articulator predicted using MFCCs and DeCoAR features for healthy controls and patients, in the fine-tuned training scheme.}
    \label{box}
\end{figure}

\vspace{-0.2cm}
Fig. \ref{box} illustrates the articulatory specific performances, for the seen cases, by DeCoAR (best SSL feature) against MFCCs. As we observe, the DeCoAR feature performs better than MFCCs for all the articulators, for both healthy controls and patients. For patients, we notice maximum relative improvements for $TT_x$ (13.58\%) and $TB_x$ (10.78\%). In the case of healthy controls, maximum relative improvements for $UL_y$ (4.62\%) and $TD_y$ (3.72\%) are achieved.

\vspace{-0.35cm}

\vspace{-0.8cm}
\section{Conclusions}
\vspace{-0.3cm}
We introduce an approach that uses pre-trained SSL features, with x-vectors, for AAI on dysarthric speech and demonstrate their effects. We perform experiments in seen and unseen subject conditions. The results revealed that, in particular, features from wav2vec, DeCoAR, and APC perform well or are on par with baseline MFCCs for dysarthric AAI in both seen and unseen conditions. We also showed that using smaller amounts of training data from an unseen target subject and trained using DeCoAR, as the input acoustic feature, achieves a superior performance owing to its unsupervised pre-training and supervised fine-tuning training scheme. Our future work includes understanding the effects of SSL networks on a language-mismatched dysarthric corpus with high severity.

\footnotesize{
\bibliographystyle{IEEEbib}
\bibliography{strings, refs}}

\begin{thebibliography}{10}

\bibitem{kiernan2011amyotrophic}
Matthew~C Kiernan, Steve Vucic, Benjamin~C Cheah, Martin~R Turner, Andrew
  Eisen, Orla Hardiman, James~R Burrell, and Margaret~C Zoing,
\newblock ``Amyotrophic lateral sclerosis,''
\newblock {\em The lancet}, vol. 377, no. 9769, pp. 942--955, 2011.

\bibitem{langmore1994physiologic}
Susan~E Langmore and Mark~E Lehman,
\newblock ``Physiologic deficits in the orofacial system underlying dysarthria
  in amyotrophic lateral sclerosis,''
\newblock {\em Journal of Speech, Language, and Hearing Research}, vol. 37, no.
  1, pp. 28--37, 1994.

\bibitem{illa2018comparison}
Aravind Illa, Deep Patel, BK~Yamini, N~Shivashankar, Preethish-Kumar Veeramani,
  Kiran Polavarapui, Saraswati Nashi, Atchayaram Nalini, Prasanta~Kumar Ghosh,
  et~al.,
\newblock ``Comparison of speech tasks for automatic classification of patients
  with amyotrophic lateral sclerosis and healthy subjects,''
\newblock in {\em 2018 IEEE International Conference on Acoustics, Speech and
  Signal Processing (ICASSP)}, 2018, pp. 6014--6018.

\bibitem{kent1991speech}
Raymond~D Kent, Robert~L Sufit, John~C Rosenbek, Jane~F Kent, Gary Weismer,
  Ruth~E Martin, and Benjamin~R Brooks,
\newblock ``Speech deterioration in amyotrophic lateral sclerosis: A case
  study,''
\newblock {\em Journal of Speech, Language, and Hearing Research}, vol. 34, no.
  6, pp. 1269--1275, 1991.

\bibitem{stein2012guide}
Cyndi Stein-Rubin and Renee Fabus,
\newblock {\em A guide to clinical assessment and professional report writing
  in speech-language pathology},
\newblock Delmar, 2012.

\bibitem{toutios2003rough}
Asterios Toutios and Konstantinos Margaritis,
\newblock ``A rough guide to the acoustic-to-articulatory inversion of
  speech,''
\newblock in {\em 6th Hellenic European Conference of Computer Mathematics and
  its Applications, HERCMA-2003}. Citeseer, 2003.

\bibitem{illa2018low}
Aravind Illa and Prasanta~Kumar Ghosh,
\newblock ``Low resource acoustic-to-articulatory inversion using
  bi-directional long short term memory.,''
\newblock in {\em Interspeech}, 2018, pp. 3122--3126.

\bibitem{maharana2021acoustic}
Sarthak~Kumar Maharana, Aravind Illa, Renuka Mannem, Yamini Belur, Preetie
  Shetty, Veeramani~Preethish Kumar, Seena Vengalil, Kiran Polavarapu, Nalini
  Atchayaram, and Prasanta~Kumar Ghosh,
\newblock ``Acoustic-to-articulatory inversion for dysarthric speech by using
  cross-corpus acoustic-articulatory data,''
\newblock in {\em 2021 IEEE International Conference on Acoustics, Speech and
  Signal Processing (ICASSP)}, 2021, pp. 6458--6462.

\bibitem{seneviratne2019multi}
Nadee Seneviratne, Ganesh Sivaraman, and Carol~Y Espy-Wilson,
\newblock ``Multi-corpus acoustic-to-articulatory speech inversion.,''
\newblock in {\em Interspeech}, 2019, pp. 859--863.

\bibitem{udupa2023improved}
Sathvik Udupa, C~Siddarth, and Prasanta~Kumar Ghosh,
\newblock ``Improved acoustic-to-articulatory inversion using representations
  from pretrained self-supervised learning models,''
\newblock in {\em 2023 IEEE International Conference on Acoustics, Speech and
  Signal Processing (ICASSP)}, 2023, pp. 1--5.

\bibitem{ghosh2010generalized}
Prasanta~Kumar Ghosh and Shrikanth Narayanan,
\newblock ``A generalized smoothness criterion for acoustic-to-articulatory
  inversion,''
\newblock {\em The Journal of the Acoustical Society of America}, vol. 128, no.
  4, pp. 2162--2172, 2010.

\bibitem{baevski2019vq}
Alexei Baevski, Steffen Schneider, and Michael Auli,
\newblock ``vq-wav2vec: Self-supervised learning of discrete speech
  representations,''
\newblock in {\em Proc. of ICLR}, 202a.

\bibitem{liu2020mockingjay}
Andy~T Liu, Shu-wen Yang, Po-Han Chi, Po-chun Hsu, and Hung-yi Lee,
\newblock ``Mockingjay: Unsupervised speech representation learning with deep
  bidirectional transformer encoders,''
\newblock in {\em 2020 IEEE International Conference on Acoustics, Speech and
  Signal Processing (ICASSP)}, 2020, pp. 6419--6423.

\bibitem{chung2019unsupervised}
Yu-An Chung, Wei-Ning Hsu, Hao Tang, and James Glass,
\newblock ``An unsupervised autoregressive model for speech representation
  learning,''
\newblock in {\em Interspeech}, 2019.

\bibitem{liu2020non}
Alexander~H Liu, Yu-An Chung, and James Glass,
\newblock ``Non-autoregressive predictive coding for learning speech
  representations from local dependencies,''
\newblock {\em arXiv preprint arXiv:2011.00406}, 2020.

\bibitem{ling2020deep}
Shaoshi Ling, Yuzong Liu, Julian Salazar, and Katrin Kirchhoff,
\newblock ``Deep contextualized acoustic representations for semi-supervised
  speech recognition,''
\newblock in {\em 2020 IEEE International Conference on Acoustics, Speech and
  Signal Processing (ICASSP)}, 2020, pp. 6429--6433.

\bibitem{schneider2019wav2vec}
Steffen Schneider, Alexei Baevski, Ronan Collobert, and Michael Auli,
\newblock ``wav2vec: Unsupervised pre-training for speech recognition,''
\newblock in {\em Interspeech}, 2019.

\bibitem{Liu_2021}
Andy~T Liu, Shang-Wen Li, and Hung-yi Lee,
\newblock ``Tera: Self-supervised learning of transformer encoder
  representation for speech,''
\newblock {\em IEEE/ACM Transactions on Audio, Speech, and Language
  Processing}, vol. 29, pp. 2351--2366, 2021.

\bibitem{illa2020impact}
Aravind Illa and Prasanta~Kumar Ghosh,
\newblock ``The impact of speaking rate on acoustic-to-articulatory
  inversion,''
\newblock {\em Computer Speech \& Language}, vol. 59, pp. 75--90, 2020.

\bibitem{illa2020closed}
Aravind Illa and Prasanta~Kumar Ghosh,
\newblock ``Closed-set speaker conditioned acoustic-to-articulatory inversion
  using bi-directional long short term memory network,''
\newblock {\em The Journal of the Acoustical Society of America}, vol. 147, no.
  2, pp. EL171--EL176, 2020.

\bibitem{snyder2018x}
David Snyder, Daniel Garcia-Romero, Gregory Sell, Daniel Povey, and Sanjeev
  Khudanpur,
\newblock ``X-vectors: Robust dnn embeddings for speaker recognition,''
\newblock in {\em 2018 IEEE international conference on acoustics, speech and
  signal processing (ICASSP)}, 2018, pp. 5329--5333.

\bibitem{rudzicz2012torgo}
Frank Rudzicz, Aravind~Kumar Namasivayam, and Talya Wolff,
\newblock ``The torgo database of acoustic and articulatory speech from
  speakers with dysarthria,''
\newblock {\em Language Resources and Evaluation}, vol. 46, pp. 523--541, 2012.

\bibitem{zue1990speech}
Victor Zue, Stephanie Seneff, and James Glass,
\newblock ``Speech database development at mit: Timit and beyond,''
\newblock {\em Speech communication}, vol. 9, no. 4, pp. 351--356, 1990.

\bibitem{kirchho1999robust}
Katrin Kirchho,
\newblock ``Robust speech recognition using articulatory information,''
\newblock {\em PhD esis, University of Bielefeld, Bielefeld, Germany}, 1999.

\bibitem{wu2015acoustic}
Zhiyong Wu, Kai Zhao, Xixin Wu, Xinyu Lan, and Helen Meng,
\newblock ``Acoustic to articulatory mapping with deep neural network,''
\newblock {\em Multimedia Tools and Applications}, vol. 74, pp. 9889--9907,
  2015.

\bibitem{udupa2021estimating}
Sathvik Udupa, Anwesha Roy, Abhayjeet Singh, Aravind Illa, and Prasanta~Kumar
  Ghosh,
\newblock ``Estimating articulatory movements in speech production with
  transformer networks,''
\newblock in {\em Proc. of Interspeech}, 2021, p. 1154–1158.

\bibitem{povey2011kaldi}
Daniel Povey, Arnab Ghoshal, Gilles Boulianne, Lukas Burget, Ondrej Glembek,
  Nagendra Goel, Mirko Hannemann, Petr Motlicek, Yanmin Qian, Petr Schwarz,
  et~al.,
\newblock ``The kaldi speech recognition toolkit,''
\newblock in {\em IEEE 2011 workshop on automatic speech recognition and
  understanding}, 2011.

\bibitem{nagrani2020voxceleb}
Arsha Nagrani, Joon~Son Chung, Weidi Xie, and Andrew Zisserman,
\newblock ``Voxceleb: Large-scale speaker verification in the wild,''
\newblock {\em Computer Speech \& Language}, vol. 60, pp. 101027, 2020.

\bibitem{van2008visualizing}
Laurens Van~der Maaten and Geoffrey Hinton,
\newblock ``Visualizing data using t-sne.,''
\newblock {\em Journal of machine learning research}, vol. 9, no. 11, 2008.

\bibitem{paszke2019pytorch}
Adam Paszke, Sam Gross, Francisco Massa, Adam Lerer, James Bradbury, Gregory
  Chanan, Trevor Killeen, Zeming Lin, Natalia Gimelshein, Luca Antiga, et~al.,
\newblock ``Pytorch: An imperative style, high-performance deep learning
  library,''
\newblock {\em Advances in neural information processing systems}, vol. 32,
  2019.

\bibitem{illa2020speaker}
Aravind Illa and Prasanta~Kumar Ghosh,
\newblock ``Speaker conditioned acoustic-to-articulatory inversion using
  x-vectors,''
\newblock in {\em Interspeech}, 2020, pp. 1376--1380.

\end{thebibliography}

\end{document}